\documentclass[conference]{IEEEtran}
\IEEEoverridecommandlockouts
\usepackage{cite}
\usepackage{hyperref}
\usepackage{amsmath,amssymb,amsfonts}
\usepackage{algorithmic}
\usepackage{graphicx}
\usepackage{textcomp}
\usepackage{siunitx}
\usepackage[version=4]{mhchem}
\usepackage{xcolor}
\def\BibTeX{{\rm B\kern-.05em{\sc i\kern-.025em b}\kern-.08em
    T\kern-.1667em\lower.7ex\hbox{E}\kern-.125emX}}

\begin{document}

\title{Multi-dimensional microwave sensing using graphene waveguides}

\author{\IEEEauthorblockN{\textbf{Patrik Gubeljak}}
\IEEEauthorblockA{Cambridge Graphene Centre\\University of Cambridge\\ Cambridge, United Kingdom}
\href{https://orcid.org/0000-0001-6955-419}{0000-0001-6955-419}\\[1.5cm]

\IEEEauthorblockN{\textbf{Luca Magagnin}}
\IEEEauthorblockA{Dipartimento di Chimica, Materiali\\e Ingegneria Chimica "Giulio Natta"\\Politecnico di Milano, Milan, Italy}
\href{https://orcid.org/0000-0001-5553-6441}{0000-0001-5553-6441}\\
\\

\IEEEauthorblockN{\textbf{Antonio Lombardo}}
\IEEEauthorblockA{London Centre for Nanotechnology\\and Department of Electronic and\\Electrical Engineering\\University College London\\London, United Kingdom}
\href{https://orcid.org/0000-0003-3088-6458}{0000-0003-3088-6458}\\
\\Corresponding author: a.lombardo@ucl.ac.uk
\and 

\IEEEauthorblockN{\textbf{Lorenzo Pedrazzetti}}
\IEEEauthorblockA{Dipartimento di Chimica, Materiali \\e Ingegneria Chimica "Giulio Natta"& \\Politecnico di Milano, Milan, Italy\\and Cambridge Graphene Centre\\University of Cambridge\\Cambridge, United Kingdom}
\href{mailto:lorenzo.pedrazzetti@polimi.it}{lorenzo.pedrazzetti@polimi.it}\\[0.2cm]

\IEEEauthorblockN{\textbf{Stephan Hofmann}}
\IEEEauthorblockA{Department of Engineering\\University of Cambridge\\Cambridge, United Kingdom}
\href{https://orcid.org/0000-0001-6375-1459}{0000-0001-6375-1459}\\

\and
\IEEEauthorblockN{\textbf{Oliver J. Burton}}
\IEEEauthorblockA{Department of Engineering\\University of Cambridge\\Cambridge, United Kingdom}
\href{https://orcid.org/0000-0002-2060-1714}{0000-0002-2060-1714}\\[1.5cm]
\IEEEauthorblockN{\textbf{George G. Malliaras}}
\IEEEauthorblockA{Department of Engineering\\University of Cambridge\\Cambridge, United Kingdom}
\href{https://orcid.org/0000-0002-4582-8501}{0000-0002-4582-8501}
}

\maketitle

\begin{abstract}
This paper presents an electrolytically gated broad-band microwave sensor where atomically-thin graphene layers are integrated into coplanar waveguides and coupled with microfluidic channels. The interaction between a solution under test and the graphene surface causes material and concentration-specific modifications of graphene’s DC and AC conductivity. Moreover, wave propagation in the waveguide is modified by the dielectric properties of materials in its close proximity via the fringe field, resulting in a combined sensing mechanism leading to an enhanced S-parameter response compared to metallic microwave sensors. The possibility of further controlling the graphene conductivity via an electrolytic gate enables a new, multi-dimensional approach merging chemical field-effect sensing and microwave measurement methods. By controlling and synchronising frequency sweeps, electrochemical gating and liquid flow in the microfluidic channel, we generate multi-dimensional datasets that enable a thorough investigation of the solution under study.  As proof of concept, we functionalise the graphene surface in order to identify specific single-stranded DNA sequences dispersed in phosphate buffered saline solution. We achieve a limit of detection concentration of $\sim 1$ attomole per litre for a perfect match DNA strand and a sensitivity of $\sim 3$ dB/decade for sub-pM concentrations. These results show that our devices represent a new and accurate metrological tool for chemical and biological sensing.
\end{abstract}


\section{Introduction}

Since the ground-breaking experiment that investigated the electronic properties of graphene in 2004\cite{novoselov2004electric}, there has been great interest in using it as a sensor material due to the large surface area to volume ratio. Many chemical species interact with it, from gases\cite{Schedin2007}, to enzymes\cite{enzyme}, to larger biomolecules\cite{Bharti2020}, and even viruses\cite{covid_graphene} and bacteria\cite{bacterium_sensor}, changing the electronic properties of graphene\cite{sensor_mechanism}. These changes can be measured directly as a change in the conductivity of the graphene FET (GFET) channel at a given operating point\cite{Zhu2016}. Another approach exploits the ambipolarity of graphene, where the charge neutrality point (CNP), i.e., the point of minimum conductivity, is tracked for different analytes or concentrations, such as in the case of the first graphene DNA sensor\cite{Hwang2016_first_DNA}. The surface properties have also been exploited to non-covalently functionalise the graphene channel, allowing high specificity without affecting the underlying electronic structure and detrimentally impacting the beneficial electronic properties. This has been used to great effect for studying the kinetics of biomaterials\cite{Xu2017_realtime_DNA} and creating highly specific sensors with record-low limits of detection\cite{Hwang2020_600zM}.

RF, micro- and mm-Waves have seen much interest in use for biosensing as they mitigate the issues seen with lower frequency measurements, such as ion screening, limited penetration and low sensitivity\cite{Artis2015} and different device architectures, ranging from reflectometers, resonators, interferometers and waveguides have been reported. Reflectometers and resonators are narrowband due to their nature, while interferometers add additional complexity in both the waveguide and microfluidic networks. For investigating the dielectric properties of biological materials, the broadband response of transmission lines such as microstrips and coplanar waveguides (CPWs) is useful and they have been used as suitable structures with some success\cite{Gonzalez2019}. However, CPW based sensors have lower sensitivity and hence limits of detection than other RF methods, so there is an ongoing search for methods to enhance sensitivity\cite{MunozEnano2019_CPW_metamaterial}.

Here we propose a novel electrochemically-gated graphene-CPW sensor coupled with a microfluid channel. Specific single-stranded DNA sequences are immobilised onto the graphene surface via non-covalent functionalisation. The sensor is then exposed to a phosphate buffered saline (PBS) solution containing complementary sequences of different concentrations. The changes in graphene's conductivity resulting from the binding (hybridisation) of complementary single-stranded DNA are measured via the variation of scattering ($S$) parameters associated with the waveguide. Furthermore, the graphene chemical potential is tuned via an electrochemical gate, providing a further dimension for the datasets generated by the sensors. We achieve a limit of detection of $\sim 1$ attomole per litre (aM) for a perfect match DNA strand and a sensitivity of $\sim 3$ dB/decade for sub-pM concentrations, showing that our sensor represents an accurate metrological tool for information-rich chemical and biological sensing.

\section{Materials and Methods}

\subsection{Design of the graphene-CPW sensor}

\begin{figure}[htbp]
\centerline{\includegraphics[width=\columnwidth]{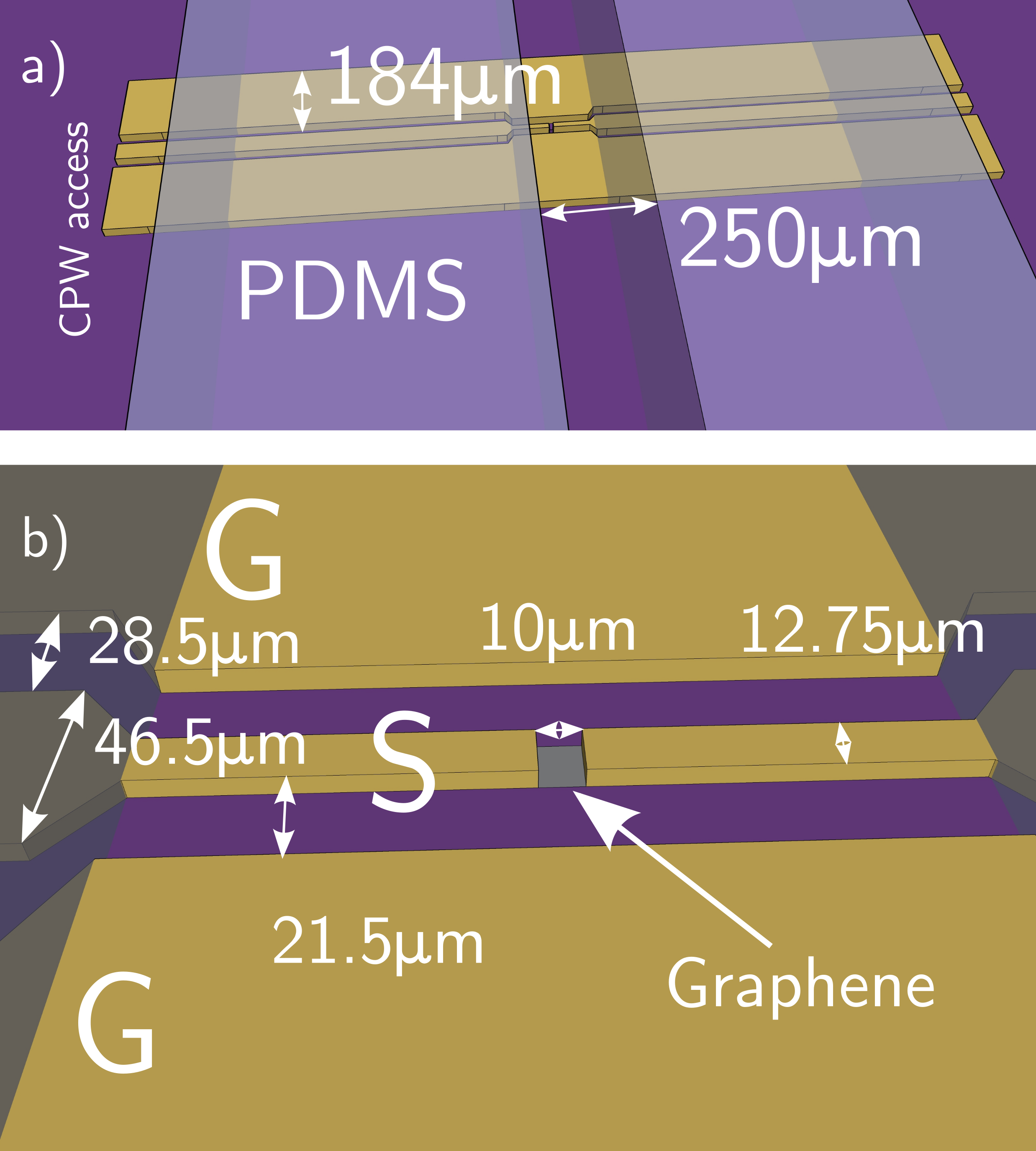}}
\caption{Sketch of the graphene-CPW sensor with relevant dimensions. The top of the enclosed PDMS channel in a) is omitted for clarity. b) shows the closeup of the tapered, graphene containing part of the CPW. S = Signal, G = Ground.}
\label{cpw_sketch}
\end{figure}

The CPW is designed to be interfaced with ground-signal-ground (GSG) probes with $150$ $\mu$m pitch. The waveguide length extends well beyond the graphene channel, as the larger surface area helps the polydimethylsiloxane (PDMS) channel adhere to the substrate better. To reduce the total number of defects in the graphene channel the area of the channel needs to be small. This is achieved by tapering the central trace of the waveguide and adjusting the gap between the G and S conductors to keep the characteristic impedance close to $50$ \si{\ohm}, as shown in Fig.~\ref{cpw_sketch}. The narrower gap also results in stronger electromagnetic (EM) field strengths, which can aid sensitivity due to the stronger interaction with the dielectric.

\subsection{Manufacturing}
The graphene was grown on copper foil according to the process described by Burton et al.\cite{Burton2019}. To transfer the graphene from the copper foil to the undoped silicon wafer with $285$ nm \ce{SiO2} the copper foil is spin-coated with polymethyl methacrylate (PMMA) 495K (8\% in anisole, MicroChem) on the graphene side. The residual carbon on the bottom side of the foil is etched away using a $10$ W \ce{O2} plasma for $60$ seconds. Then the copper foil is floated on a $0.75\%$ wt solution of \ce{(NH4)2S2O8} overnight. Once the copper is etched, the floating PMMA-graphene sheet is collected by a cleaned glass plate and floated on the surface of deionised water for 10 minutes and then again in fresh DI water for another 10 minutes to remove any residual persulfate. The Si wafer is cleaned by consecutive sonication in acetone and isopropanol, after which it is exposed to $10$ W \ce{O2} plasma for 60 seconds to remove any organic contaminants. The PMMA-graphene sheet is then picked up using the cleaned Si substrate and left to dry overnight. Once dry, the Si wafer is immersed in acetone overnight to remove the PMMA. A cleaning immersion for $10$ minutes first in fresh acetone and then in fresh isopropanol is also performed to remove as much of the PMMA as possible. The wafer with graphene on top is then annealed for $5$ minutes at $150$ \si{\celsius}.

The graphene channel is patterned using AZ$5214$E photoresist and direct-write laser lithography and etched using $3$ \si{\W} \ce{O2} plasma for $30$ seconds. Another layer of AZ$5214$E is applied and the metal contacts patterned using the above method. Before metal deposition the graphene-metal contact areas are exposed to $0.5$ \si{\W} \ce{Ar} plasma for $20$ seconds to improve contact resistance. $5$ nm of \ce{Cr} and $100$ nm of \ce{Au} is deposited using a Kurt J. Lesker e-beam evaporator, with liftoff in acetone. The resulting CPWs are then encapsulated using a PDMS channel. The fluid inlet and outlet are punched out, feedlines introduced and sealed to the mould using liquid PDMS. The finished device assembly is then cured at $80$ \si{\celsius} overnight to bond the channel to the wafer and cure the seals.

The completed fluid assembly is then used to functionalise the graphene with the probe single DNA strand (P20), as reported by Campos et al.\cite{Campos2019_attomolar_DNA}. After the passivation with ethanolamine, the device is ready for DNA concentration measurements.

\subsection{Measurements}

\begin{figure}[htbp]
\centerline{\includegraphics[width=\columnwidth]{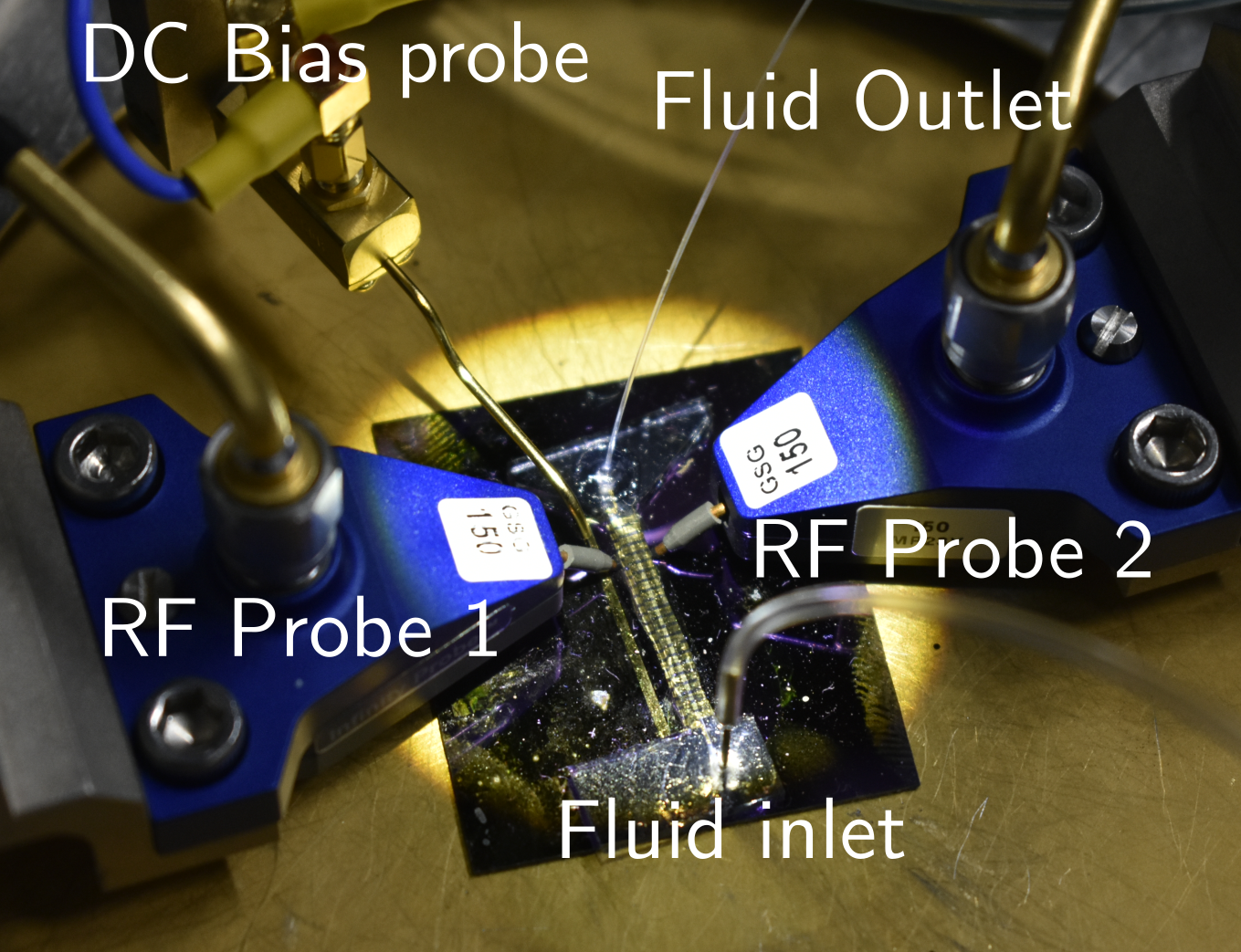}}
\caption{The prepared microfluidic, DC and RF setup. The device under test consist of an array of graphene CPWs sharing the same microfluidic channel.}
\label{RF_setup}
\end{figure}

Solutions for functionalisation and measurement are introduced to the graphene devices using an ElveFlow OB1 MK3 microfluidic controller with a flow meter. All measurements are performed in $0.01X$ phosphate buffered saline (PBS) solution, at a flow rate of $2$ \si{\micro \liter \per \minute}. The sample is mounted on a Cascade Microtech Summit 1200B semi-automatic probe station, connected to an Agilent B1500A Parameter Analyser and PNA-X 5245 vector network analyzer (VNA). Connection to the CPW is established using a pair of Cascade Microtech i50 GSG 150 Infinity Probes, connected via bias networks to the parameter analyser. The RF setup is calibrated using an impedance standard substrate with the mTRL method. The RF power was set to $-12$ dBm, set to scan between $50$ MHz and $50$ GHz to collect $1001$ data points, with an intermediate frequency (IF) of $1$ kHz. Electrolytic gating is applied by the parameter analyser using a Cascade Microtech DCP100 probe contacting the RF ``cage", which also acts as the DC gate electrode, immersed in the solution. The ready-to-measure setup is shown in Fig.\ref{RF_setup}.

To measure the response of the sensor to different DNA concentrations, the desired concentration of the target perfect match DNA (T20) is flowed through the channel at $2$ \si{\micro \liter \per \minute} for $1$ hour and then rinsed with $0.01X$ PBS for 5 minutes. We perform $10$ consecutive DC transfer curve sweeps to saturate the hysteretic behaviour, after which the $S$-parameters are measured. To facilitate large scale measurements, we developed an automated control software which performs DC measurements, applies the gate bias and measures the S-parameters for all gate voltages of interest.

\section{Results}

\subsection{S-parameter measurements with no gate voltage applied}
\begin{figure}[htbp]
\centerline{\includegraphics[width=\columnwidth]{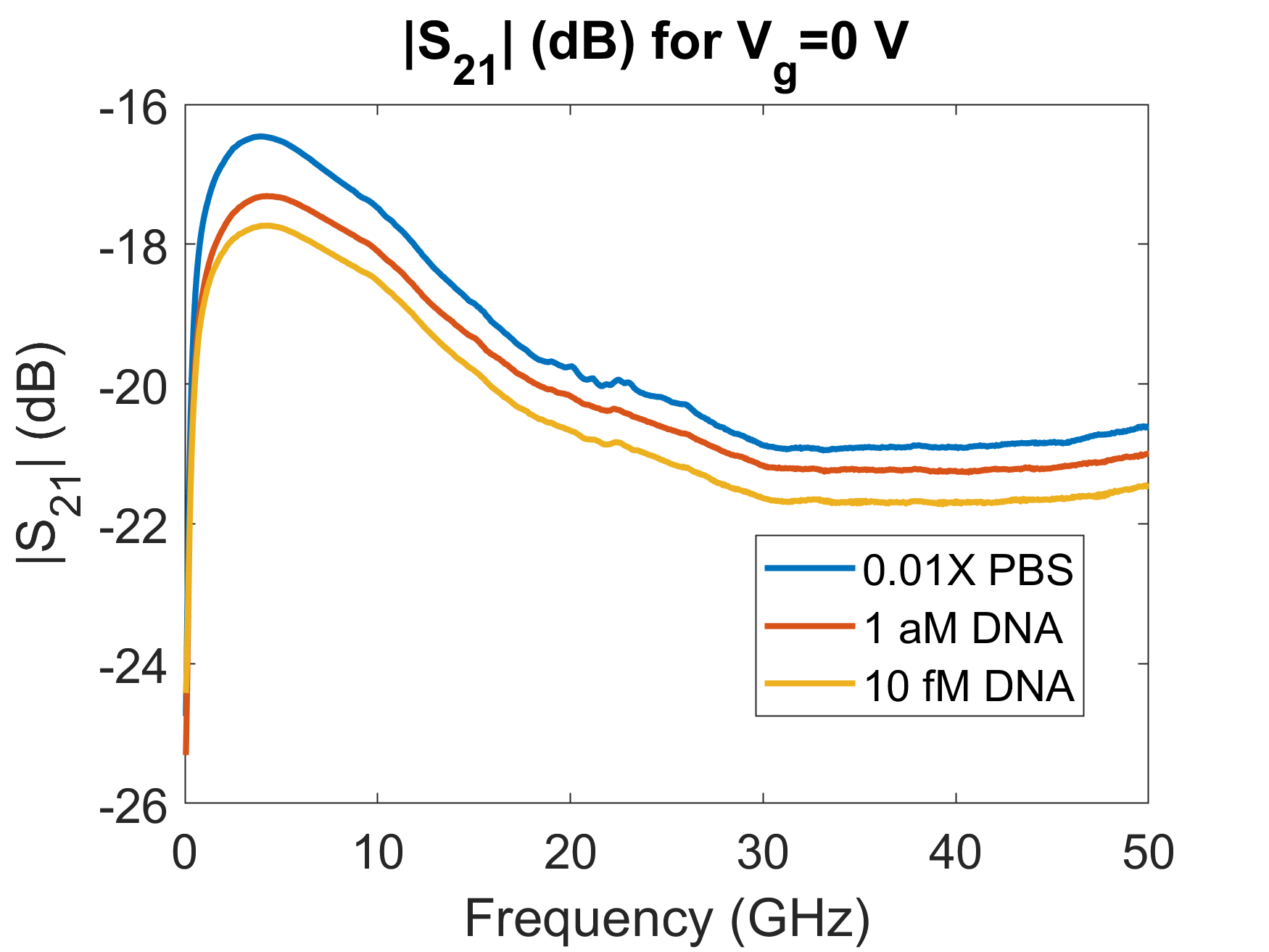}}
\caption{The $|S_{21}|$ parameters at $V_g=0$ V, mimicking standard all metal CPW sensors.}
\label{S_at_0Vg}
\end{figure}

First, we tested our sensor without applying any gate voltage. We sequentially expose the surface of the graphene CPW to PBS and solutions containing DNA at different concentrations. Fig.~\ref{S_at_0Vg} reports the magnitude of the $S_{21}$ parameter at different DNA concentrations, showing the ability of our graphene CPW sensor to detect hybridisation of our probe with a perfect match target DNA to concentrations as low as $1$ aM.

In order to compare our results with metallic waveguides used for dielectric spectroscopy we also include metallic CPWs with identical dimensions. We expose them to the same concentration of DNA, however we do not observe any change in scattering parameters. This is expected, as -unlike graphene- the metal conductivity is not altered by the DNA and the functionalisation protocol used is unlikely to immobilise DNA onto the metal surface. 

\subsection{Multi-dimensional approach}

\begin{figure}[htbp]
\centerline{\includegraphics[width=\columnwidth]{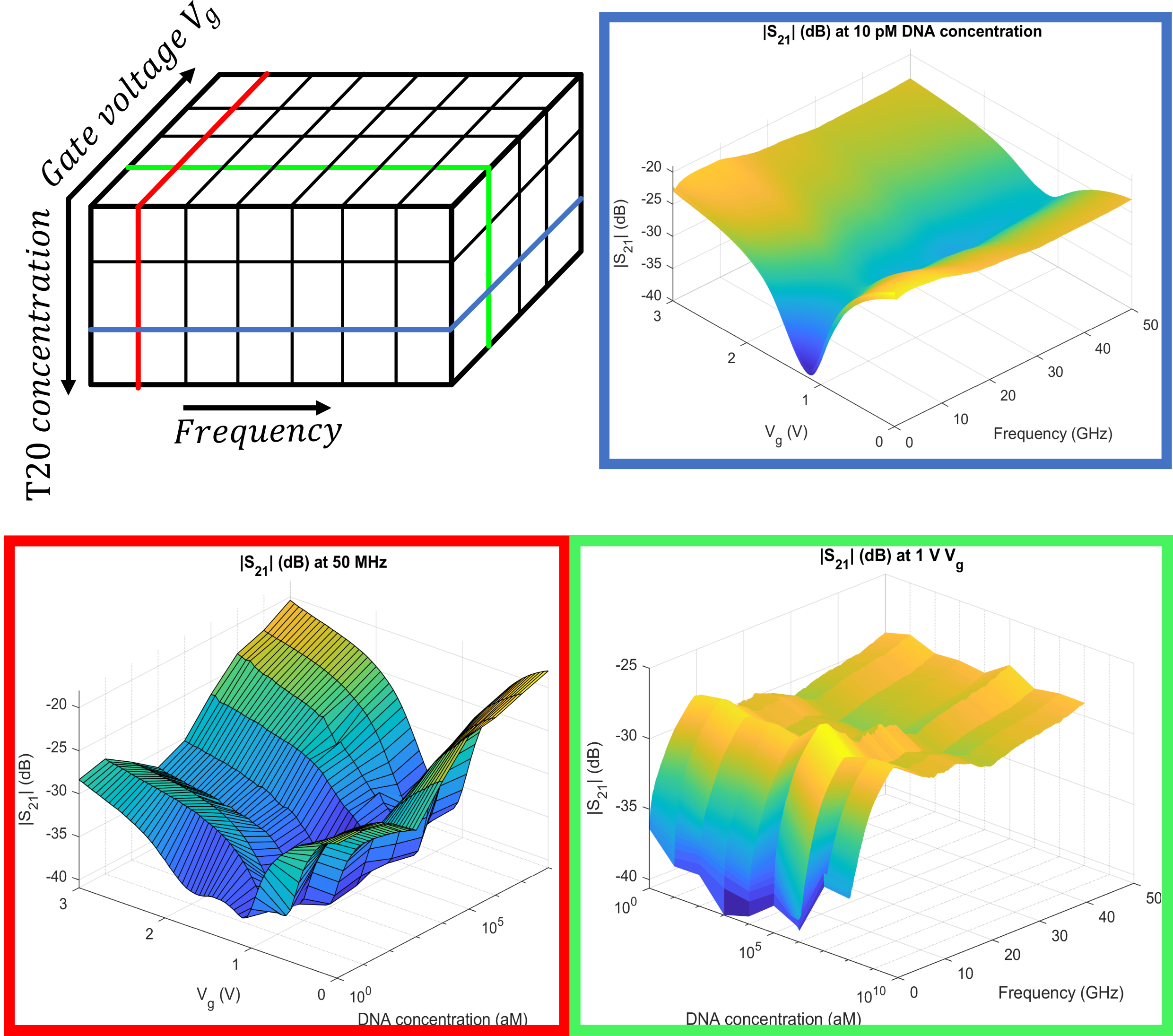}}
\caption{Illustration of the multidimensional nature of the dataset.}
\label{multi_data_set}
\end{figure}

Graphene conductivity can also be controlled by a suitable gate, enabling a novel approach combining sensing achieved via Chemical FETs at DC and dielectric spectroscopy at microwave frequencies, effectively creating a higher-dimensional data set. As such, the techniques of comparing transfer curves or $S$-parameters for different concentrations or species, are not able to fully take advantage of the available data. We achieve this by measuring the $S$-parameters for a large frequency range from $50$ MHz to $50$ GHz, limited by our equipment (the biasing network and the VNA). For a given chemical concentration, we measure the $S$-parameter frequency sweep, change the gate voltage and measure a new $S$-parameter sweep. With reference to Fig~\ref{multi_data_set}, each dataset (generated in the form of a $.S2P$  file) represents one line of cubes along the frequency axis. The next line is added to its side, in the direction of the gate voltage axis, which consists of the $.S2P$ measurements at the same concentration, but a different gate voltage. Performing this for all gate voltages creates one page, i.e., the complete gate voltage dependent $S$-parameters for a given concentration. Then, once the concentration is changed, the procedure is repeated, and the new page is stacked on top of the existing one, along the concentration axis of the now Rank 3 Tensor data structure.

Forming the dataset in such a way allows for new approaches to understanding the effect of the analyte on $S$-parameters. Slicing the dataset along different axes gives different surfaces which can be used independently or in conjunction to see the changes in the $S$-parameters from introducing the analyte, shown for the example of $|S_{21}|$ in Fig.~\ref{multi_data_set}. Similarly, depending on the device, $S_{11}$ could be a more relevant parameter for resonance based devices, and the surfaces of the real and imaginary components can be compared. For CPW devices, we are more interested in the transmission of the EM waves, so the $S_{21}$ parameter is of more interest, as it contains more information on the dielectric properties of the surrounding dielectric, i.e., the sample under test. Moreover, the $S$-parameters are complex, so the dimensionality is even higher, due to each $S$-parameter actually representing a two-component vector ($[Re(S_{xy})\quad Im(S_{xy})]$ or $[|S_{xy}|\quad \angle S_{xy}]$).

\subsection{Sensitivity from data surfaces}

\begin{figure}[htbp]
\centerline{\includegraphics[width=\columnwidth]{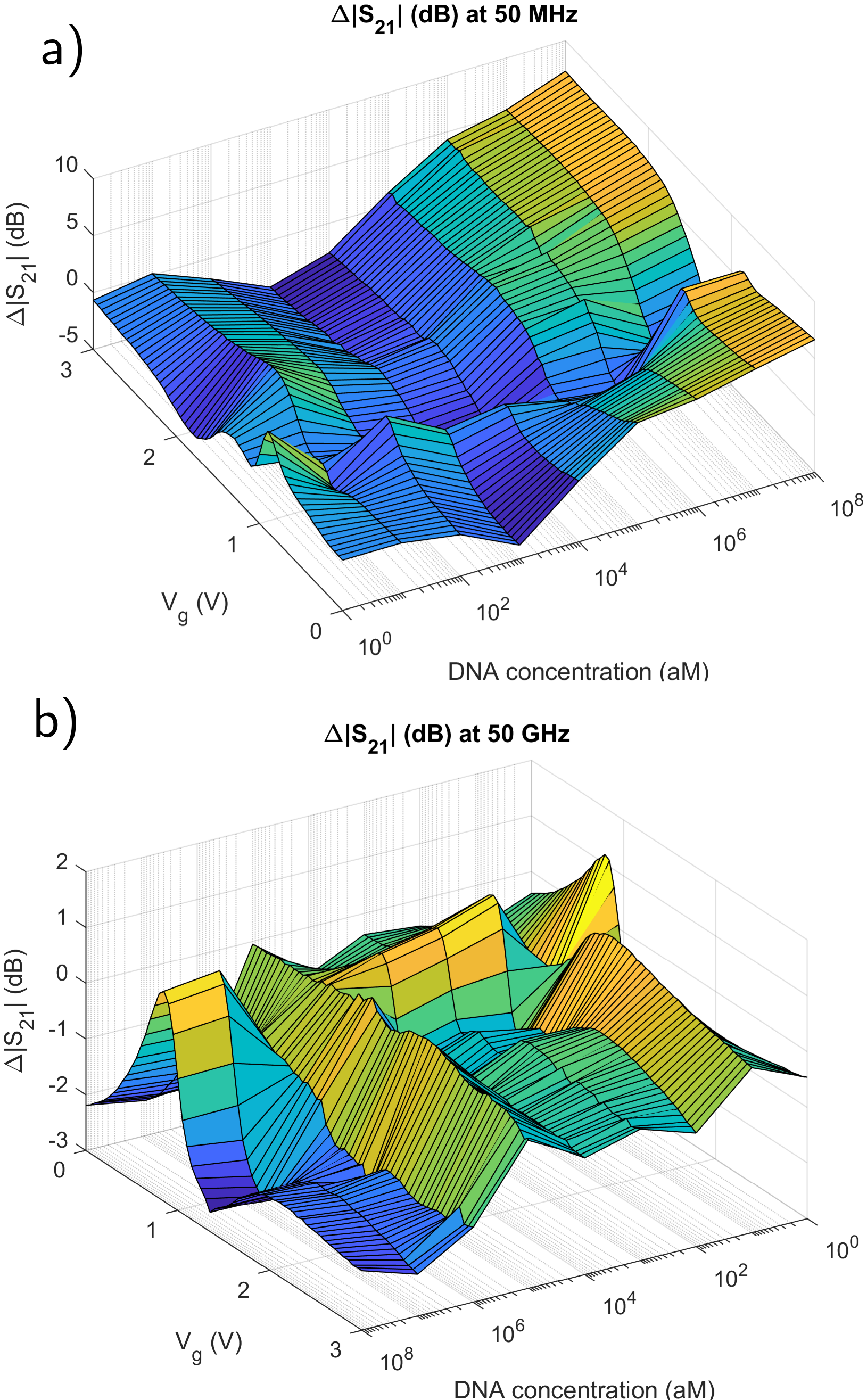}}
\caption{The $\Delta|S_{21}|$ surfaces at $50$ MHz and $50$ GHz. The orientations of the concentration and gate voltages axes in the figures are opposite for easier presentation of differences in surface features.}
\label{dS21_50MHz}
\end{figure}

In order to analyse the data obtained we consider only the absolute value of $S_{21}$ and how it changes with DNA concentration and gate voltage at specific frequencies, which is  equivalent to taking the red slice in the 3D dataset shown in Fig.~\ref{multi_data_set}. We define $\Delta|S_{21}|$ surfaces by subtracting, at each gate voltage, the $|S_{21}|$ value corresponding to the $0.01X$PBS solution without any target DNA, i.e., we identify variations of $|S_{21}|$ in the DNA data with respect to to the zero concentration data at the same gate voltage. Fig.~\ref{dS21_50MHz} (a) and (b) show two representative $\Delta|S_{21}|$, corresponding to frequencies of $50$ MHz and $50$ GHz respectively. 

Such surfaces can be constructed for every measured frequency, leading to two-dimensional data sets. A thorough analysis would require considering the entire surface and extracting all the relevant features, however, it is possible to simplify the analysis by identifying lines of maximum change along the concentration axis at a given gate voltage, from which sensitivity and limit of detection of the sensor can be estimated. By measuring the changes for the trend at $V_g=3$~V in Fig.~\ref{dS21_50MHz} (a), we obtain a sensitivity of $2.85$ dB/decade in the range from $1$ fM to $100$ pM. As shown in Fig.~\ref{dS21_50MHz}(b), the $\Delta|S_{21}|$ surface becomes more complex at higher frequencies, with more peaks and valleys and would be better analysed as whole, e.g. by applying feature extraction algorithms. The analysis of such complex surfaces is beyond the scope of this work; however, we speculate that the simultaneous analysis of multiple features could result in higher sensitivity and lower limit of detection.  

\subsection{Surface-to-surface comparison for sensing}

\begin{figure}[htbp]
\centerline{\includegraphics[width=\columnwidth]{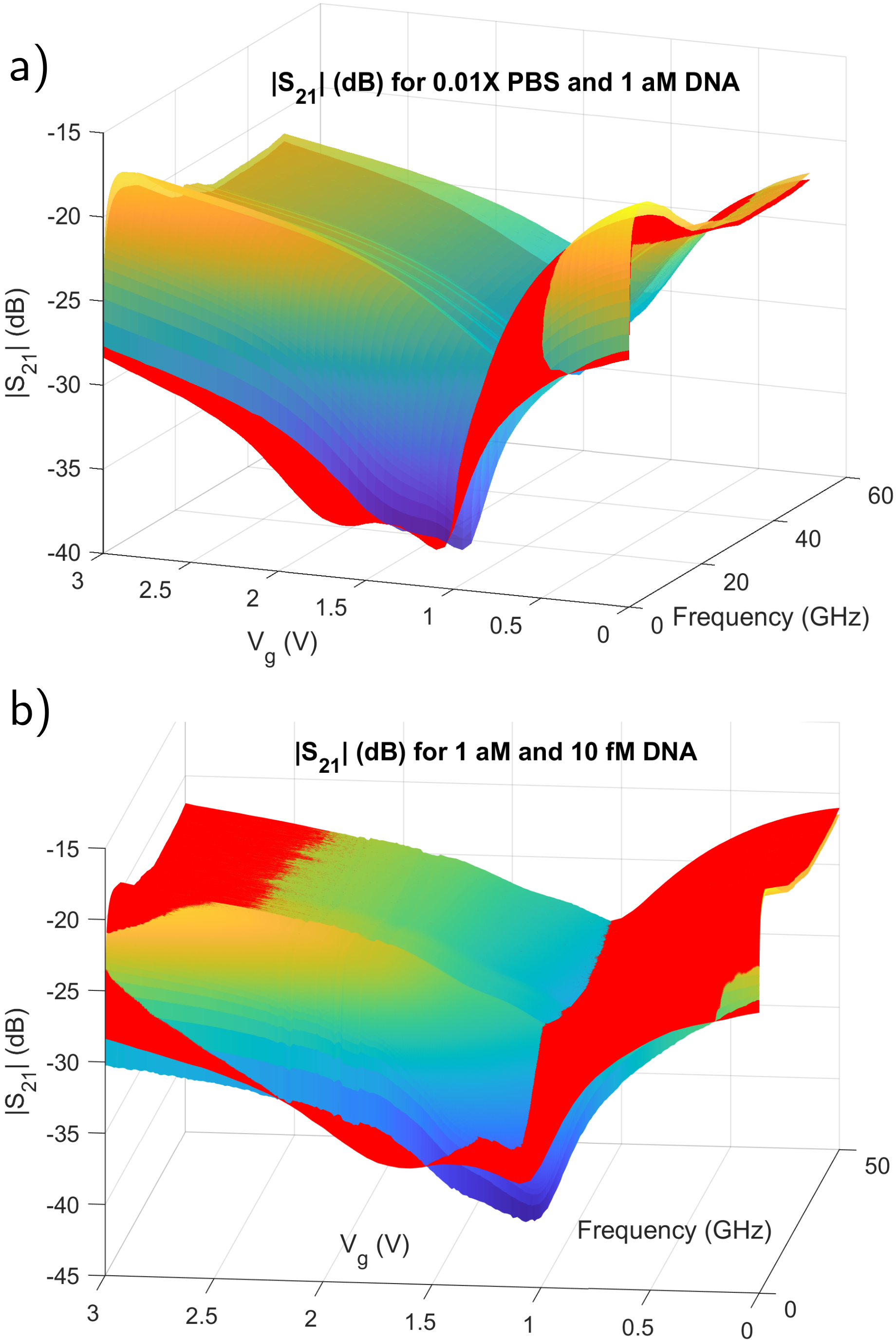}}
\caption{a) The surfaces of $|S_{21}|$ for $0.01X$PBS (colour gradient) and $1$ aM T20 (red). b) The surfaces of $|S_{21}|$ for $10$ fM T20 (colour gradient) and $1$ aM T20 (red).}
\label{fig:PBS_vs_1aM}
\end{figure}

The detection of low analyte concentrations can be done in a similar way by comparing the Frequency$-V_g$ surfaces of $0.01X$PBS, $1$~aM DNA and $10$ fM DNA, shown in Fig~\ref{fig:PBS_vs_1aM}.

It can be readily seen that there is a clear difference in the response of the sensor for the different analyte concentrations. However, that difference is not easily quantifiable; for certain operating conditions, i.e., gate voltages, the trend of the changing $S$-parameter magnitude is opposite. If the response was solely due to the change in chemical potential and the resulting conductivity, the surfaces would only be shifted w.r.t. each other along the $V_g$ axis, while having the same shape. However, as the introduction of the analyte, i.e., DNA strand and the resulting hybridisation, changes the dielectric environment as well, the propagation of high frequency EM waves is affected in more subtle ways. These manifest themselves as the changes in the maxima, minima and gradients of the surfaces. 
These features all contain information on the dielectric behaviour of the analyte at different electrochemical conditions, with high sensitivity. Multiple features can increase the reliability of detection, provide evidence for the presence of analytes with sufficient confidence to lower the limits of detection and allow for wider characterisation of dielectric properties of biomaterials.

\section{Conclusions}

By incorporating graphene into a coplanar waveguide coupled with microfluidic channels we achieve a new type of sensor, unifying (electro-)chemical and high frequency sensing principles. Demonstration of the sensor using DNA hybridisation as the test (bio-)analyte shows state-of-the-art limits of detection, while extending the operating range by orders of magnitude compared to conventional DC approaches. Our sensors produce multi-dimensional datasets and represent a novel metrological tool for chemistry and biology. 


\bibliographystyle{ieeetr}
\bibliography{bibliography}

\vspace{5mm} 
\textbf{Acknowledgements}: We acknowledge funding from EPSRC grant EP/L016087/1

\end{document}